# Constraining age and metallicity of bulges embedded in low surface brightness galaxies

G. Galaz*, L. Infante*, A. Villalobos† and J. Dalcanton**

*Departamento de Astronomía y Astrofísica, Pontificia Universidad Católica de Chile, Casilla 306, Santiago 22, Santiago, Chile
†Kapteyn Astronomical Institute, Postbus 800 9700 AV Groningen, The Netherlands
**Department of Astronomy, University of Washington, Box 351580, Seattle, WA 98195, USA

**Abstract.** We present results from deep Magellan and VLT long slit optical spectroscopy for a sample of 19 face-on low surface brightness galaxies (LSBGs) selected from the sample of Galaz et al. (2002). In particular, we compute Lick/IDS indices for the light emitted by the bulges hosted by these galaxies. Similar procedure is performed for a sample of 14 face-on high surface brightness sample from the sample of Peletier & Balcells (1996). Results indicate that bulges of LSBGs are metal poorer, compared to those hosted by HSBGs, supporting the hypothesis that stellar formation rates have been stable low during the lifetime of these galaxies. We compare our results with those of Bergmann, Jorgensen & Hill (2003).

## BACKGROUND

Low surface brightness galaxies (LSBGs) have been the subject of interest since Disney [8] emphasized that the fact that the central surface of disk galaxies in the Hubble sequence Sa-Sb-Sc fall in a rather narrow range, might be the result of a selection effect which is due to the difficulty in discovering galaxies of very low surface brightness. In practice, LSB means galaxies whose central surface brightness is fainter than 22.0 mag arcsec$^{-2}$ in the $B$ band, and seem to have different properties that those of brighter galaxies, which might imply a different evolutionary track. It was originally thought [18] that all galaxies with low surface brightness were early or late-type dwarfs. However, radial velocity observations [9] showed that some LSBGs are actually quite large and luminous [13, 4].

Previous studies [12] showed that LSBGs are much more gas rich than high surface brightness ones and bluer than "normal" late type galaxies [13, 4, 16]. These facts combined with the measured low metallicity [14, 3], result in a low star formation rate (SFR) in unevolved system. However, much of the information which has been obtained, and serve as base for further analysis, comes from the analysis of broadband photometry data. Given the low surface brightness of the underlying population, few spectroscopic endeavors have been successful to obtain a more precise picture of the stellar populations of LSBGs [6, 1], in particular for their fossil stellar content (equivalent widths, abundances, etc.). Only recently, with the advent of 8m class telescopes, it has been possible to measure the metallicity LSBGs of their old stellar populations.

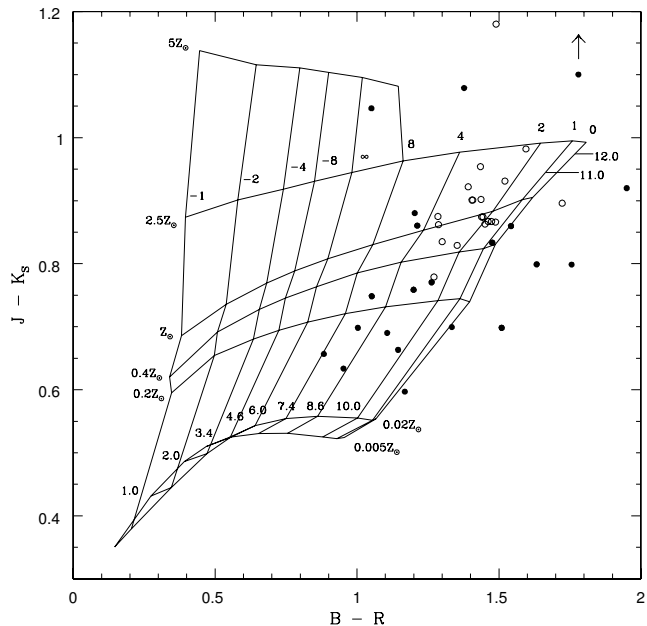

**FIGURE 1.** Optical and near-IR colors for bulges hosted by our sample of face-on LSBGs (filled dots), and for our sample of bulges in face-on HSBGs (these from [15]). The grid indicates the isochrones and locus of equal metallicity, using the models of Bruzual & Charlot (2003) [5]. Note the lower $J - K_s$ color for the bulges hosted by LSBGs, suggesting a lower metallicity.

## OBSERVATIONS AND ANALYSIS

In this proceeding, we present preliminary results of a study on stellar populations in LSBGs. In particular, we are interested in the old, fossil stellar population, which is present in the bulges of these galaxies. We present results from long slit spectroscopy of a selected subsample from Galaz et al. [11]. The subsample is composed by 19 face-on spiral galaxies with $21.7 < \mu_{0,disk}(B) < 23.4$ mag arcsec$^{-2}$. Most of the galaxies have HI masses between $10^{9.5} M_\odot$ and $10^{9.9} M_\odot$. Therefore, they are massive LSBGs. For these galaxies, we have already measured $B$, $R$, $J$ and $K_s$ broadband colors, suggesting that these bulges are metal poor, as presented in Figure 1 from [10]. Also, most of them are face-on galaxies. This constraint prevent large internal extinction and ensures that the light entering the long slit comes mainly from the bulge of the galaxy. Furthermore, for most galaxies, we were able to observe part of the disk and spiral arms, allowing to obtain emission line spectra[1] for the stellar formation regions. Along with the 19 LSBGs, we observed 14 face-on high surface brightness (HSB) galaxies from the sample of Peletier & Balcells [15]. For the long slit spectroscopic observations we use the 6.5m

---

[1] In many cases, however, emission lines appear simultaneously in absorption (stellar phase, from the nucleus) and emission (gas phase, from the disk, spiral arms). We have used the spectrophotometric models to correct this effect, in a similar way as explained in [1].

Magellan telescope and the VLT, with the Boller & Chivens and FORS1 spectrographs, respectively, between April and October 2002 and 2003. For some galaxies (4) we also used the Magellan LDSS2 and IMACS spectrograph. Total exposure times varied between 50 minutes and 2.3 hours, depending on the central surface brightness. Signal to noise ratios were in general no less that 30, in the range 4500 - 5500 Å. Each spectrum was extracted, wavelength and flux calibrated, following standard routines for long slit reduction. Typical spectral effective resolution was 3 - 6 Å.

## Lick/IDS indices for bulge stellar populations

Several Lick/IDS indices [17] were calculated using absorption lines. In particular, we compute absorption indices H$\beta$ (bandpass 4847 - 4876 Å), Mgb (5160 - 5192 Å) and <Fe> = (Fe5270 + Fe5335)/2 (Å), for both the LSB and the HSB galaxy sample. For this purpose we use both a manual method (using IRAF package) and the code AUTOINDEX, this last kindly provided by Nicolás Cardiel [7]. Special caution was taken to compute (1) the continuum corresponding for each index, and (2) the absorption line fit in the cases where H$\beta$ has absorption *and* emission features. When we deal with this kind of feature, we use a similar method presented in Bergmann, Jorgensen & Hill (2003) [1] (BJH2003 hereafter), but using the new models of Bruzual & Charlot (2003) ([5]). Results are shown in Figure 2, where we also compare with the work of BJH2003. The grids in Figure 2 shows equal age (isochrones) after an instantaneous burst, and equal metallicity, as indicated by labels. It is apparent from Figure 2 that (1) thanks to the high S/N ratio ($\sim 30 - 60$) of our spectra, it is possible to obtain smaller error bars, compared to those presented for the LSBGs observed by BJH2003, obtaining also a more restricted range of indices values by the subsequent scatter, (2) the stellar populations embedded in bulges of LSBGs present similar H$\beta$ index compared to that hosted by HSBGs bulges, *but smaller* Mgb and <Fe> indices, indicating that they are metal poorer systems, compared to the stellar populations hosted by bulges of HSBGs.

This lower metallicity agree with the observational fact that in the overall, LSBGs are metal poor compared to HSBGs, at least from result obtained by several authors from metallicity measurements from the gas phase [2] using emission lines, which in turn agree (or is obtained from) with low stellar formation rates (SFRs). A lower metallicity measured from the stellar phase using absorption lines, in particular from those stellar populations lying in the nucleus, could indicate that LSBGs have experimented a low SFR during their whole lifetime, not only during the recent past.

## ACKNOWLEDGMENTS

GG thanks Fondecyt grant #1040359. LI thanks FONDAP #15010003 "Center for Astrophysics". Authors would like to thank the organizers of the workshop for the excellent welcome they bring to all the participants.

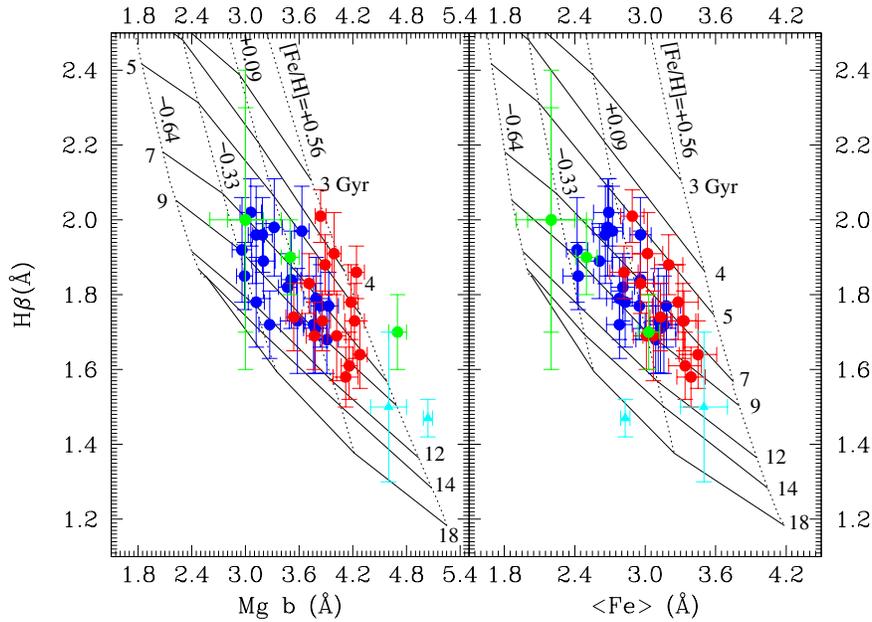

**FIGURE 2. Left panel**: Measured Lick/IDS indices Mgb (5160-5192 Å) and H$\beta$ (4847-4876 Å) for our sample of bulges hosted by 19 face-on LSBGs (blue dots), compared with the same indices measured for a sample of bulges in 14 face-on high surface brightness galaxies (HSBGs) from the sample of Peletier & Balcells (1996) [15] (red dots). Other symbols show results from Bergmann, Jorgensen & Hill (2003) [1], superimposed to this plot. In green, results for their LSBGs sample with measured indices, and in cyan, results for their HSBGs sample. The grid shows lines of equal metallicity (expressed in fractions [Fe/H], dotted lines) and equal age (solid lines), obtained from the spectrophotometric models of Bruzual & Charlot (2003) [5]. The age means the age after an instantaneous burst of stellar formation. **Right panel**: The same as left panel but for the Lick/IDS indices H$\beta$ and <Fe> = (Fe5270 + Fe5335)/2 (Å).

# REFERENCES

1. Bergmann, M., Jorgensen, I., & Hill, G. 2003, AJ, 125, 116 (BJH2003)
2. de Blok, W., & van der Hulst, J. 1998, A&A, 335, 421
3. de Blok, W., van der Hulst, J., & Bothun, G. 1995, MNRAS, 274, 235
4. Bothun, G., Impey, C., & McGaugh, S. 1997, PASP, 109, 745
5. Bruzual, G., & Charlot, S. 2003, MNRAS, 344, 1000
6. Burkholder, V., Impey, C., & Sprayberry, D. 2001, AJ, 122, 2318
7. Cardiel, N., Gorgas, J., Cenarro, J., & González, J. 1998, A&AS, 127, 597
8. Disney, M. 1976, Nature, 263, 573
9. Fisher, J., & Tully, R. 1975, A&A, 44, 151
10. Galaz, G., Villalobos, A., & Infante, L. 2004, in preparation
11. Galaz, G., Dalcanton, J., Infante, L., & Treister, E. 2002, AJ, 124, 1360
12. Longmore, A., Hawarden, T., Goss, W., Mebold, U., & Webster, B. 1982, MNRAS, 200, 325
13. McGaugh, S., Schombert, J., & Bothun, G. 1995, AJ, 109, 2019
14. McGaugh, S. 1994, ApJ, 426, 135
15. Peletier, R., & Balcells, M. 1996, AJ, 111, 6
16. Sprayberry, D., Bernstein, G., Impey, C., & Bothun, G. 1995, ApJ, 438, 72
17. Trager, S., Worthey, G., Faber, S., Burstein, D., & González, J. 1998, ApJS, 116, 1
18. van den Bergh, S. 1959, AJ, 64, 347